# Scattering Amplitude of $\pi N$ on the Second Sheet

**Mohamed E. Kelabi**

Physics Department, Faculty of Science, University of Tripoli,
Tripoli, Libya

**Abstract**

We propose parameterization procedures of the scattering amplitude $f_{1+}^{3}(s)$ with a view to extracting the pole parameters from data in the elastic region of $\pi N$ scattering. This is achieved by considering the analyticity properties of partial wave amplitudes directly and writing dispersion relations for the amplitude explicitly.

**Introduction**

The continuation of scattering amplitudes through two-particles branch lines has been considered by authors [1], [2]. It has also been stated in the analyticity hypothesis of Mandelstam [3], [4], [5], that the complete amplitude for scattering may be continued as an analytic function in the complex energy plan with the exception of certain regions of singularities [6], shown in Figure 1.

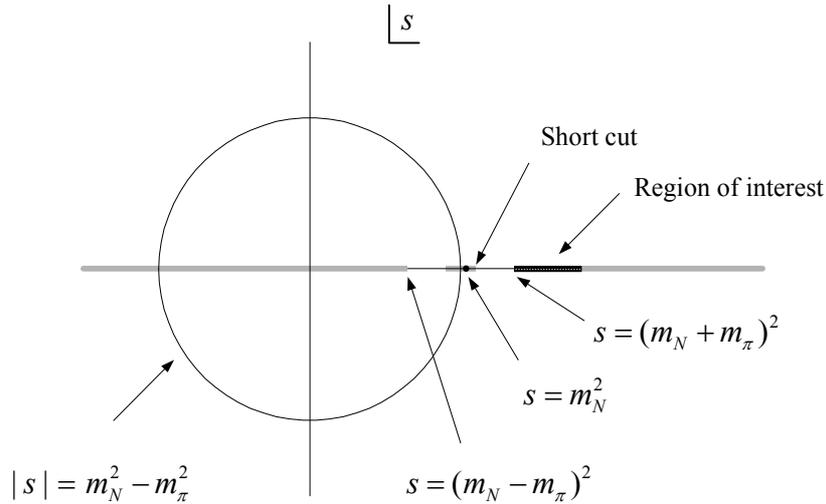

Figure 1. Singularities in the partial wave amplitudes for pion-nucleon scattering. Figure adapted from [6].

The partial wave amplitude $f(s)$ for the scattering nucleon-pion of masses $m_N$, $m_\pi$ may be continued through the elastic cut $(m_N + m_\pi)^2 < s \leq s_i$, where the kinematic variable $s_i$ [7] indicates the threshold for inelastic process. The continuation on the second sheet is given by the relation [1], [8]

$$f^{II}(s) = \frac{f(s)}{1 + 2i\, q(s) f(s)} \qquad (1.1)$$

with

$$q(s) = \frac{\left[\left(s - (m_N + m_\pi)^2\right)\left(s - (m_N - m_\pi)^2\right)\right]^{1/2}}{2 s^{1/2}}$$

which has the properties [8], [9]

$$q^*(s^*) = -q(s)$$

$$q(s + i0) \geq 0$$

for real $s \geq (m_N + m_\pi)^2$. The continuation function $f^{II}(s)$ has poles which do not appear in $f(s)$ correspond to the vanishing denominator in Eq. (1.1), the resonance poles $(1 + 2i\, qf)$. It has been suggested [1] that the poles in $f^{II}$ may be taken into account by continuing the contour integration which give the dispersion relation through the two-particle branch cut and thus obtaining a two sheeted dispersion relation. If we write dispersion relation for the difference

$$\tilde{f}(s) = \frac{1}{2}[f(s) - f^{II}(s)]$$

along the real axis in the elastic region $(m_N + m_\pi)^2 < s \leq s_i$, this gives

$$\mathrm{Re} f(s) = \frac{1}{\pi} \int_L ds' \frac{\Delta \tilde{f}(s')}{s' - s} + \frac{\mathcal{P}}{\pi} \int_{s_{in}}^{\infty} ds' \frac{\mathrm{Im} \tilde{f}(s')}{s' - s}$$

$$+ \sum_n \left( \frac{r_n}{z_n - s} + \frac{r_n^*}{z_n^* - s} \right) \qquad (1.2)$$

where $\Delta \tilde{f}(s')$ are the discontinuities of the left hand cuts, $z_n$ are any complex resonance poles, and $r_n$ are proportional to their complex residues.

**Parameterization of the Scattering Amplitude**
In what follows we wish to propose parameterization procedures based on Eq. (1.2) for the scattering amplitude $f(s) = f_{1+}^3(s)$. To do this we assume that the inelastic cut $s_{in} \leq s \leq \infty$ is very weak and can be safely neglected, and that only one pair of resonance poles $s = \alpha \pm i\beta$ with residues $r = a \pm ib$ need to be retained, that is

$$\mathrm{Re}\, f(s) = \frac{1}{\pi}\int_L ds' \frac{\Delta \tilde{f}(s')}{s' - s} + \frac{a + ib}{\alpha + i\beta - s} + \frac{a - ib}{\alpha - i\beta - s} \qquad (1.3)$$

We also parameterize the left-hand cuts by simple poles, which will be chosen to be a pole at $s = m_N^2$ to represent the Born cut, and another pole at $s = s_2 = 20$, which is approximately at the center of the crossed-cut for $\Delta(1232)$ exchange. Finally, we exploit the property $\tilde{f}(s_0) = 0$ at threshold to write a subtracted dispersion relation. In this way we obtain a simple parameterization of the form

$$f(s) = U(s) - U(s_0) \qquad (1.4)$$

where

$$U(s) = \frac{c_1}{s - m_N^2} + \frac{c_2}{s - s_2} + \left( \frac{a + ib}{\alpha + i\beta - s} + \frac{a - ib}{\alpha - i\beta - s} \right)$$

contains six free real parameters[*)]: $c_1$, $c_2$, $a$, $b$, $\alpha$, and $\beta$, which can be fitted to data to give the positions of the resonance poles $\alpha \pm i\beta$ and their complex couplings $a \pm ib$.

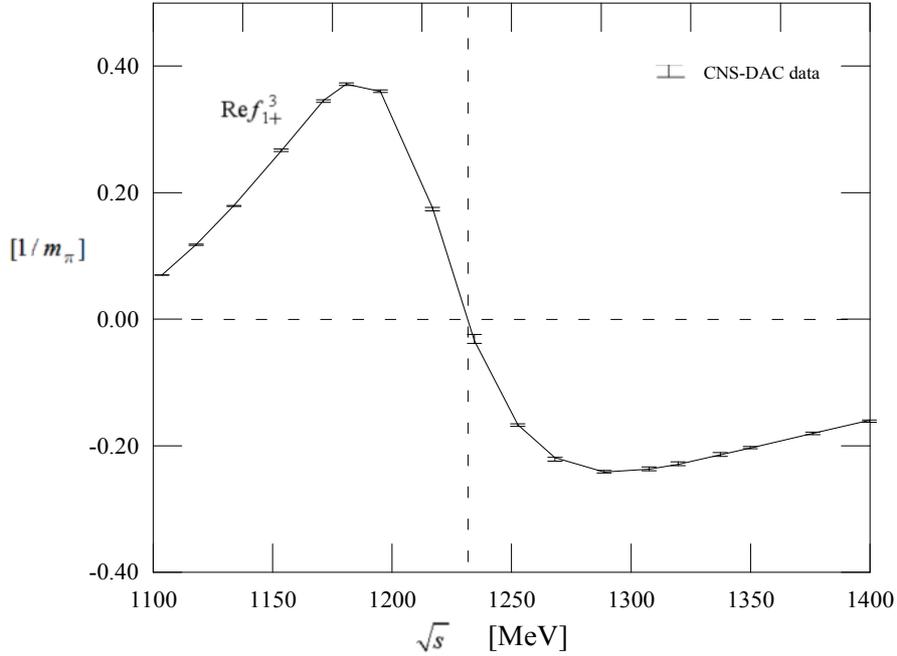

Figure 2. The Scattering amplitude for $p\pi^+$ in the resonance region.

---

[*)] All values are in the power of the charged pion mass units.

In Figure 2, we show the results of our fitting based on Eq. (1.4) for $p\pi^+$ in comparison with data from CNS-DAC [10]. The fit yields the following chi-squared per degrees of freedom

$$\chi^2 = 2.3 \times 10^{-6} \quad [1/m_\pi]$$

giving the fit parameters in charged pion mass units

$$c_1 = 1.48005 \pm 0.32495$$
$$c_2 = -20.70486 \pm 1.52489$$
$$a = 1.71242 \pm 0.01608$$
$$b = 1.26665 \pm 0.02084$$
$$\alpha = 75.13127 \pm 0.03256$$
$$\beta = 6.22344 \pm 0.03939$$

with the corresponding pole positions

$$W_r \pm i\frac{\Gamma}{2} = (1210.80 \pm 0.20) \pm i(50.06 \pm 0.09) \quad [\text{MeV}]$$

these values can be compared with earlier results listed in Table 1. The value at which the real part of scattering amplitude passes through the zero i.e., the phase shit passes through $90^0$, is

$$\tilde{W}_r = 1231.5 \quad [\text{MeV}]$$

| Real Part | $-2 \times$ Imaginary Part | Name | Year |
| --- | --- | --- | --- |
| 1210.8±0.2 | 100.1±0.2 | Present work | 2013 |
| 1210.8 | 99 | Höhler [11] | 2001 |
| 1211±1 to 1212±1 | 102 ±2 to 99±2 | Hanstein [12] | 1996 |
| 1206.9±0.9 to 1210.5±1.8 | 111.2±2.0 to 116.6±2.2 | Miroshnichenko [13] | 1979 |
| 1208.0±2.0 | 106±4 | Campbell [14] | 1976 |

Table 1. Pole positions comparison of different fits for the $\Delta(1232)$ resonance.

**Conclusion**
We see there is a reasonable consistency between the resonance pole parameters obtained by different methods for $\pi N$ elastic scattering, this is because the resonance dominates the relevant partial waves. In the case where background is large, more care may be required. In such cases, we think the method described here may be more reliable, since it incorporates the resonance poles directly, in a rather model independent way, while remaining flexible and easy to use.

**Acknowledgement**

I would like to thank the Physics department at the University of Malta for their invitation and warm hospitality extended to me.